%% file: paper.tex
\documentclass[aps,preprint,groupedaddress,showpacs]{revtex4}

\input symdefs

\def\qqquad{\qquad\qquad}

\def\sumsp{\sum_{{\rm sp}}}
\def\dL{d\Lambda}
\def\dW{d{\cal W}}
\def\dV{d{\cal V}}

\def\intW{\int\dW\,}

\def\scriptl{{\cal L}}

\def\scriptp{{\cal P}}

\def\avg#1{\Bigg\langle #1\Bigg\rangle}

\def\Astar{\vec A^*}
\def\Bstar{\vec B^*}
\def\Bpl{B_\parallel^*}
\def\Bpp{B_\perp}

\def\pMM#1{{\pt #1\over\pt M}}

\def\zpl{z_\parallel}
\def\zperp{\vec{z_\perp}}

\def\xdot{\vec{\dot x}}

\def\Rdot{\vec{\dot R}}
\def\thetadot{\dot\vartheta}
\def\Mdot{\dot M}

\def\pzdot{\dot\zpl}

\def\zdot{\dot z}
\def\Zdot{\dot Z}
\def\zpdot{\vec{\dot Z_p}}
\def\pzzp#1{{\pt #1\over\pt\vec Z_p}}
\def\pvpvp#1{{\pt #1\over\pt\varphi}}
\def\ppzpz#1{{\pt #1\over\pt\zpl}}

\def\Epsilon{{\cal E}}
\def\Epslash{{\cal E}\hskip -0.2 cm / \hskip 0.05 cm}
\def\epslash{\epsilon\hskip -0.15 cm / \hskip 0.001 cm}

\def\Apol{\vec A_{{\rm pol}}}
\def\Btor{\vec B_{{\rm tor}}}

\def\ppp{p_\perp}
\def\Ppl{P_\parallel}
\def\Mphi{M_\varphi}
\def\Pphi{\scriptp_\varphi}
\def\bphi{b_\varphi}
\def\Piphi{\vec\Pi_\varphi}
\def\rhom{\rho_m}
\def\rhogc{\rho_G}
\def\Jgc{\vec J_G}
\def\vor{\varpi}

\begin{document}
\title[Gyrokinetic Gauge Transform]
{Gyrokinetic Field Theory as a Gauge Transform\\ 
or: gyrokinetic theory without Lie transforms}
\author{Bruce D. Scott}
\email[email: ]{bds@ipp.mpg.de}
\affiliation{Max-Planck-Institut f\"ur Plasmaphysik, 
		Boltzmannstr 2,
                D-85748 Garching, Germany}

\date{\today}

\begin{abstract}
Gyrokinetic theory is a basis for treating magnetised plasma
dynamics slower than particle
gyrofrequencies where the scale of the background is larger than
relevant gyroradii.  The energy of field perturbations can be comparable
to the thermal energy but smaller than the energy of the background
magnetic field.  Properly applied, it is a low-frequency gauge transform
rather than a treatment of particle orbits, and more a 
representation in terms of gyrocenters rather than particles than an
approximation.  By making all transformations and approximations in the
field/particle Lagrangian one preserves exact energetic consistency so
that time symmetry ensures energy conservation and spatial axisymmetry
ensures toroidal angular momentum conservation.  This method draws on
earlier experience with drift-kinetic models while showing the
independence of gyrokinetic representation from particularities of Lie
transforms or specific ordering limits, and that the essentials of
low-frequency magnetohydrodynamics, including the equilibrium, are
recovered.  It gives a useful basis for total-f electromagnetic
gyrokinetic or gyrofluid computation.
Various versions of the representation based
upon choice of parallel velocity space coordinate are illustrated.
\end{abstract}


\maketitle

\section{Background and Introduction
\label{sec:intro}}

The most modern form of gyrokinetic theory appeared about two decades ago
in two papers by Sugama and Brizard \cite{Sugama00,Brizard00}.  The
gyrokinetic approach to low frequency motion had emerged before 1980
first as an ordering scheme \cite{Rutherford68,Taylor68}, and then as
a method to average fast time scales out of the collisionless
Boltzmann equation describing evolution of a distribution function
accounting for motion of charged particles in the presence of
prescribed or self consistent electromagnetic fields governed by
Maxwell's equations \cite{Catto77,Catto78,Catto81,FriemanChen82}.  The
key to self consistency was a method to re-cast charge density in the
form of a gyrocenter charge density and a polarisation density, which
allowed solving for a low frequency electrostatic potential in the
absence of finite explicit charge density.  The low frequency
approximation in this context is equivalent to quasineutrality: the
divergences of the current and magnetic potential should vanish and
the actual charge density should vanish even in the presence of finite
ExB vorticity, whose existence implies a finite divergence of the
perpendicular electric field.  This was a matter of using the existing
approximations to form a gyrocenter representation and derive a
gyrokinetic Poisson equation \cite{Lee83}.  About the same time the
theory given a stronger mathematical foundation by demonstrating that
the original results could be recovered by applying a Lie transform to
the Lagrangian of the charge particles, so that all ordering
assumptions could be collected into the starting point of the theory.
The equations then descend rigorously from the
Euler-Lagrange equations, following earliest variational
treatments of drift-kinetic motion
\cite{Morozov66}, first for drift centers and then for
finite-gyroradius gyrocenters
\cite{Littlejohn81,Littlejohn82,Littlejohn83,
CaryLittlejohn83,WhiteChance84}.
The Lie transform contains an
opposite pull-back transformation which allows systematic derivation
of the gyrocenter representation and thereby the gyrokinetic Poisson
equation \cite{CaryLittlejohn83,Dubin83}.  The strategy of maintaining
``canonical representation'' in the Lagrangian by systematic
application of the transform's gauge freedom within a generally
covariant version of the theory was explicitly established, and then
several forms were presented under various levels of approximation
\cite{Hahm88,Brizard89,Brizard95,Hahm96}.  This includes a version
which was explicitly electromagnetic \cite{Hahm88a}.  In an important
concurrent line the representation of magnetohydrodynamics (MHD) by
gyrokinetic theory was explicitly established
\cite{Naitou95,Qin99,Qin00}.  The low-frequency form of MHD, called
Reduced MHD, restricts the dynamics to eliminate the ``fast wave''
\cite{Strauss76,Strauss77}.  The form of MHD most relevant to tokamak
dynamics adds ``low-beta'' restrictions to this ($\beta=2\mu_0
p/B^2\ll 1$), and is captured by gyrokinetic theory by allowing
fluctuations in the magnetic potential only parallel to the
background magnetic 
field (i.e., $\ptb\vec A=\Afl\vec B/B$).  Demonstrations
of such ``shear-Alfv\'en'' dynamics were given by Lee et al
\cite{Lee01,Lee03}.  These methods can also be described by a
geometric viewpoint \cite{CaryLittlejohn83,Qin04}.  A fully
relativistic, electromagnetic treatment considering the representation
of the complete Maxwell equations and exact conservation laws was
given by Brizard \cite{Brizard99}.  This paper was the central
precursor to the field theory papers referenced above
\cite{Sugama00,Brizard00}.  These have different emphases, both with
regard to relativistic or mostly low-frequency forms, and continuum or
particle representations.  Fully equivalent, they allow choice in the
approach to the theory.  Both explicitly use the Noether theorem to
obtain the conservation laws which therefore follow rigorously once
the appropriate choice of Lagrangian has been made, and it is a
particle/field system Lagrangian, not just a particle one --- this is
the step which turns the gyrokinetic representation into a field
theory.  The use of quasineutrality itself is no longer arbitrary; it
follows directly from the assumption in the system Lagrangian that the
electric field energy is small compared to the ExB kinetic energy of
the particles, and then the Euler-Lagrange equations for both
particles and fields maintain exact consistency.  The above has been
comprehensively reviewed by Brizard and Hahm \cite{Brizard07}.
Important demonstrations contained there are that the pullback
transform and variational method to obtain the fields are
mathematically identical, and that the Lie transform for the
Lagrangian and Poisson bracket transform for the kinetic equation
yield identical results, even if the field theory methods make the
connection to the rest of physics that much clearer.  However, the
field theory methods have the advantage of restricting ordering
assumptions to the starting point with no loss of consistency, whereas
ordering applied directly to the equations without regard to the
consistency of the starting point most often leads to a breach of some
or all of the conservation laws.  In a physical situation where
ingredients with small energy content can have large effect (\eg,
flows \cite{Waltz94,ZLin98,Diamond05}, or parallel currents
\cite{HasWak83,WakHas84,dalfloc,gem}, in pressure driven dynamics), it is
imperative to maintain exact consistency in the equations in any
computational model.

Energetic consistency refers not only to the existence of an exact
conservation law for some quantity definable as energy within the model,
but more generally to the principles of physical symmetry by which
fields and particles interact with each other in the dynamics within the
model.  Conserved energy and momenta are 
specific quantities derived within the model along with its
equations for the evolution of the fields and particles.  The archetype
for this is a Lagrangian system, with free pieces for each constituent
and interaction pieces describing their exchanges, such as described in
the text by Landau and Lifshitz \cite{LandauLifshitz}.  
The Euler-Lagrange equations for the particle positions give
their evolution, and the Euler-Lagrange equations for the field
potentials give the field equations.  Application of Liouville's theorem
to the particle equations gives the evolution of the particle
distribution function 
(see, e.g., pp 48--52 of the text by Tolman \cite{Tolman}).
In each application, these
equations are specific to the particular model (Lagrangian); that is,
there is no external field equation one can appropriate, but each
model has its own field equation arising from its own Lagrangian;
otherwise, energetic consistency will be broken.
Any version of the gyrokinetic model is consistent as long as all
approximations are made in constructing the Lagrangian while the
derivation of the resulting equations for the particles and fields
remains exact.  In
this context, gyrokinetic theory is a low-frequency gauge transformation
of the Maxwell-Boltzmann system Lagrangian, not an approximate ``orbit
averaging'' done on the equations themselves (this equivalence is
present only in the first-order linearised version of the gyrokinetic
representation). 

Symmetry of interaction (Newton's Third Law) is an automatic
feature of any such system,
and conservation laws are described by application of Noether's
theorem.  In gyrokinetic theory applied to tokamaks or other
axisymmetric equilibria, time symmetry leads to energy
conservation and axisymmetry leads to toroidal momentum conservation,
not just for particles moving in prescribed fields but generally for the
field/particle system.  The gyrokinetic field theory papers introduced
these applications to our context \cite{Sugama00,Brizard00}, and a
detailed exposition of why and how it works, including discussion of the
importance of canonical representation in the Lagrangian, is given in a
previous work on energetic consistency and momentum conservation in
gyrokinetic field theory \cite{momcons}.  In this context, canonical
representation refers to the strategy of the Lie transform as discussed by
Hahm \cite{Hahm88} to arrange all dependence upon space-time dependent
field quantities into the time component of the Lagrangian, so that the
canonical momenta, dependent upon phase space coordinates, are
independent of both time and toroidal angle.  The derivations described
below are arranged from the start to follow this structure, so as to
automatically guarantee the existence of energy and toroidal momentum
conservation laws, that is, general energetic consistency.

\section{Outline of Gyrokinetic Theory as a Gauge Transform
\label{sec:outline}}

Gyrokinetic theory is not an orbit average over equations, but a set of
operations on a Lagrangian which involves a change of representation
from particle to gyrocenter variables.  What those gyrocenter variables
actually are is the result of choices made during the gauge
transformation.  A gauge transformation is a combination of operations
involving coordinate changes and the addition of total
differentials to the original Lagrangian to produce another one which
reflects the same dynamics in different language as the original one.
The low-frequency approximations enter through a chosen ordering scheme,
in which the only really essential element is the smallness of the ExB
vorticity or other dynamical frequencies compared to gyrofrequencies.
Since it is the slowest, the ion gyrofrequency sets this limit.
Since they are the fastest, the electron and/or shear Alfv\'en parallel
transit
frequencies are considered.  The ExB vorticity is considered since it
underlies any turbulent dynamics involving ExB motion.  The
approximations involved in these are well satisfied in tokamaks, usually
by at least two orders of magnitude, even in steep gradient regimes.
The only significant exception is the borderline case of the outboard
midplane in present-day spherical tokamaks where the magnetic field
strength drops to relatively small values and the gradient scale length
drops to below a centimeter (e.g., \cite{Maingi12}).
For L-H transition databases on conventional tokamaks, however,
this frequency ordering is well satisfied
\cite{Suttrop97,Groebner01}.

The procedure we will use closely follows Littlejohn's variational
method from 1983 \cite{Littlejohn83}, in combination with the field
theoretical methods from Landau and Lifshitz \cite{LandauLifshitz}.
The dynamical role of the field potentials is taken on equal footing to
the gyrocenter motion, treating the field/particle system as a whole.
No separation between equilibrium or dynamics except for the
backgorund magnetic field is prescribed, expecting the electromagnetic
version of the theory to recover not only equilibrium flows but also
the MHD (Grad-Shafranov) equilibrium self-consistently.

We strictly maintain the original gyrokinetic strategy to
preserve canonical representation by transforming field variable quantities
strictly into the time component of the system Lagrangian (whether as
part of the Hamiltonian or of the field Lagrangian density).  In
Landau-Lifshitz terms we have the free-particle Lagrangian (in our terms
the part not dependent upon dynamical fields), the interaction
Lagrangian (the part involving both fields and particles), and the
free-field Lagrangian.  Canonical representation
refers to the interaction Lagrangian appearing only in the Hamiltonian
such that the canonical momenta involve only the phase space
coordinates.  All terms due to the field potentials appear only 
in $H$, so that the phase space Jacobian is a background quantity,
symmetry of the background is not broken.  This allows easily realisable
versions of the relevant conservation law proofs as well as facilitating 
proof of correspondence to conventional models \cite{momcons}.
It also facilitates computations.

The usual assumptions of low-frequency theory are quasineutrality (the
neglect of space-charge effects while allowing a finite Laplacian of the
field potential upon which finite vorticity depends), and shear-Alfv\'en
magnetic responses.  These are effected within the theory by neglecting
the $\eps_0 E^2/2$ electric field energy in the field Lagrangian and by
allowing only a parallel component of the magnetic potential $\Apl$
among magnetic disturbances.  Specifically, neglecting $\eps_0 E^2/2$
against ExB kinetic energy $\rhom(E/B)^2/2$ is a statement that
the (mostly ion) plasma polarisability $\rhom/B^2$ overshadows
the permittivity of free space $\eps_0$, so that the overall charge
density is neglected despite the nonzero divergence of the electric
field.  Since $\eps_0\mu_0=1/c^2$ this also means that the Alfv\'en
velocity $v_A$ is small compared to $c$, the speed of light: $v_A^2$
is $B^2/\mu_0 \rhom$ so that $\eps_0 B^2/\rhom$ is also $v_A^2/c^2$,
the more familiar measure.  The second statement is the neglect of
$\vec A_\perp$ disturbances due to $\beta\ll 1$ and $\kpp
v_A\gg\omega$ being well satisfied, with $\omega$ tracking $\ppt{}$ in
any time-dependent response. 
Hence in force balance $\ddpp(2\mu_0 p + B^2)\approx 0$ the
changes to the field strength $B^2$ due to the pressure $p$ are
neglected, and in the dynamics only the parallel electric field should
involve inductive responses.  Dynamical magnetic compressibility is
therefore disallowed.  However, the magnetic compressibility implied by
the existence of diamagnetic flows and heat fluxes and polarisation
currents is explicitly kept in the theory, such that any low-beta
compressibility effects are retained.

Ultimately, however, gyrokinetic theory is about the representation, not
the ordering.  We have gyrokinetic polarisation (Poisson) and induction
(Amp\`ere) equations in 
which a polarisation density appears, while the gyrokinetic equation
itself has no $\ppt{}$ terms associated with polarisation drifts.  The
polarisation current is recovered by taking the time derivative of the
polarisation equation and the MHD Ohm's law is recovered 
by taking the time derivative of the induction equation.  These two
statements recover nonlinear low-frequency MHD, including the
Grad-Shafranov equilibrium.  The representation
is the result of having preserved canonical representation.  With this
maintained the result of conventional gyrokinetic approaches is
easily recovered in the appropriate limit.

The Lie transform which corresponds to this gauge transform version was
given previously \cite{Miyato09},
including the correspondence to low-frequency MHD and to previously
derived forms of the gyrokinetic Lagrangian and equations. Generally,
energetic consistency and momentum conservation, including the route
back to MHD, follows easily in this formulation \cite{momcons}.

\section{Gauge Transform to get the Lagrangian}

Herein we derive the gyrokinetic Lagrangian as a field theory.
Finite ExB Mach number (flow amplitude) is allowed by taking a maximal
ordering on the ratio of kinetic to thermal energy (i.e., they appear at
the same level in the expansion).
The theory is cast as a gauge
transform which does not require Lie transform techniques.
The method closely follows
Littlejohn's differential gauge transform method \cite{Littlejohn83},
just that we generalise the role of the field potentials to become
dependent variables, enforce canonical representation on $L$ including them,
and treat the result as a field theory rather than a Lagrangian for
individual gyrocenters.

We consider generally a particle Lagrangian $L_p$ which gets transformed
to a gyrocenter one.  The structure is 
\begin{equation}
L_p\,dt = \vec p\cdot\vec{dZ} - H\,dt
\end{equation}
cast as a fundamental one-form,
where the components of $\vec p$ are canonical momenta, 
the components of $\vec Z$ are the phase space coordinates,
and $H$ is the Hamiltonian (the time component).  
In general this is six-dimensional (6D) dynamics, but in its
gyrokinetic representation
the gyromotion involving perpendicular velocity space components is
separated away so that the actual dynamics covers the 4D space of
$(\vec R,\zpl)$ consisting of gyrocenter positions and the parallel
canonical momentum.  Collisions bring in the 5th coordinate, usually the
magnetic moment $\mu$ conserved by the drift motion.  The sixth
coordinate is the gyrophase angle $\vartheta$ which only appears in the
gyromotion since $H$ and $\vec p$ and the rest of $L_p$ is gyrophase
independent. 

Starting with the Landau-Lifshitz treatment as a background
\cite{LandauLifshitz}, we restrict to non-relativistic situations with
the time not being varied.  We have the free particle and interaction
Lagrangians as
\begin{equation}
L_p = {m\over 2}\xdot\cdot\xdot + e\vec A\cdot\xdot - e\phi
\end{equation}
for the particles, and the free field piece $\scriptl_f$ which we treat later.
The Legendre transform is applied as
\begin{equation}
\vec p \equiv \pt L_p/\pt\xdot
\qquad
H \equiv \vec p\cdot\xdot - L_p
\qquad
L_p = \vec p\cdot\xdot - H
\end{equation}
after which $H$ and therefore $L_p$ are functions of $(\vec x,\vec p)$
and not $(\vec x,\vec v)$.  Then, $L_p$ is turned into a fundamental
one-form by considering the differential action
\begin{equation}
L_p\, dt = \vec p\cdot\vec{dx} - H\,dt
\qquad\hbox{where}\qquad
H = m{U^2\over 2} + e\phi\qquad
m\vec U = \vec p - e\vec A
\end{equation}
Since we will use $\vec A$ as an anchor for low-frequency drift motion
we split the canonical momentum into an equilibrium part and a
dynamical part,
\begin{equation}
L_p\, dt = \vec p\cdot\vec{dx} - H\,dt
\qquad\hbox{where}\qquad
\vec p = e\vec A+\vec z
\quad\hbox{and}\quad
H = m{v^2\over 2} + e\phi
\end{equation}
This gets us to the structure referred to in the beginning, but now
with $\vec p$ as total canonical momentum and $\vec z$, the dynamical
part, as velocity coordinate.  Note that $\vec x$ is used in place of
$\vec Z$ at this point because all the canonical momenta are spatial
when we start.

To get a low-frequency low-beta kinetic Lagrangian we assume $\phi$ is a
dynamical field but that $\vec A$ evolves through small shear-Alfv\'en
disturbances parallel to $\vec B$.  We re-cast $\vec A$ generally in
terms of an equilibrium piece $\vec A$ assumed to be static, and add to
it the dynamical piece $\Apl\bunit$ assuming $\Apl$ to be the other
dynamical field.  Now, $\vec A$ and $\vec B=\curl\vec A$, with
$B=\abs{\vec B}$ and $\bunit=\vec B/B$,
are assumed to be static quantities describing
the background magnetic field, while the particle coordinates and $\phi$
and $\Apl$ constitute the set of dependent variables which
represent the dynamical system.  We assume the gyromotion is a fast
frequency to be eliminated, retaining dynamics on the time scale of the
ExB vorticity and the parallel transit frequencies.  This leads to $m/e$
as the formal small parameter for expansion (which tracks the ratio
$\Omega_{exb}/\Omega_i$ between the ExB vorticity and ion gyrofrequency).

We are using $\vec A$ to anchor drifts, so we cannot use canonical
variables directly.  The gyrokinetic Lagrangian therefore represents a
non-canonical transformation \cite{Zaitsev69}.
However, we do want canonical
representation, which means that the whole Lagrangian except for $H$ is
static (dependent on geometry, coordinates, and constants only) and, in
a tokamak, axisymmetric.  It also means that the resulting phase-space
Jacobian retains the symmetry of the background, and that there are no
extra $\ppt{}$ terms on fields in the kinetic equation.  We get
canonical representation using the gauge freedom of the transformation:
generating functions of the coordinate changes, and the gauge terms
(pure differentials) added to the Lagrangian.  All time- (and in a
tokamak, toroidal angle-) dependence involving fields is moved into $H$
and out of the part of $L_p$ containing the canonical momenta.

We expand the spatial
canonical momentum in terms of parallel and perpendicular
motion, reckoned against the background magnetic field,
\begin{equation}
\vec z = \zpl\bunit+\zperp
\end{equation}
At this point, $\zpl$ contains $\Apl$ and $\zperp$ can contain a mean
flow, denoted $\vec u_0$,
as well as the gyromotion velocity.  These appear in the
Lagrangian, which is
\begin{equation}
L_p\, dt = \LP e\vec A+\zpl\bunit + \zperp\RP\cdot\vec{dx} - H\,dt
\end{equation}
\begin{equation}
H = m{U^2\over 2} + {1\over 2m}\abs{\zperp}^2 + e\phi
\qquad mU = \zpl-e\Apl
\end{equation}
where $U$, now just the parallel velocity,
is not a coordinate but a function of dependent variables.
Since there is no dynamical $\Aperp$
kept in the low-beta shear-Alfv\'en case, $\zperp$ is the same as
$m\vperp$, which may or may not involve $\vec u_0$.
We will demand any $\vec u_0$ not be specified but
emerge naturally from the derivation process.

The next step most closely follows Littlejohn's drift-kinetic gauge treatment
 \cite{Littlejohn83}.  The small parameter is formally any factor of
$m/e$ which tracks drifts.  Flows (gradients of $\phi$) will enter
 naturally, later.  A drift-kinetic representation will treat the gyromotion
$\vec w$ 
but leave a mean flow $\vec u_0$ as part of $\vec p$
whose $\ppt{}$ 
represents the polarisation drift.  In a gyrokinetic representation we
treat $\vec w$ and $\vec u_0$ together in $\zperp$, with any dependence
upon $\phi$ absent from $\vec p$ but present in $H$.  
Polarisation no longer enters as a drift, but as a density in the field
equation for $\phi$ which we will see later.  As we will show, the same
expression for $\div\vec J=0$ is recovered after $\ppt{}$ is taken on
this field equation.  We emphasise that ``gyrokinetic'' refers to the
representation, not the finite-gyroradius (FLR) effects, and that
a model with zero-FLR and polarisation density is still a gyrokinetic
one.  This will become obvious only when the self consistent field equations
are at hand.

We introduce an arbitrary spatial coordinate change in which 
\begin{equation}
\vec x = \vec R + \vec r
\end{equation}
where we can choose $\vec r$ according to how we want to arrange the
representation.  We expect its magnitude to satisfy $r\ll L_B$ where
$L_B$ is the scale of variation of the magnetic field (assumed to be
of order the toroidal major radius in a
tokamak).  We then Taylor-expand $\vec A$
and $\phi$ in powers of $\vec A$ and arrange them by order.  This
ultimately leads to a long-wavelength version of the model since we
will find {\it a posteriori} that $\rs^2\ddpp$ should be small (its
terms arise at second order).  The one-form $L_p\,dt$ splits according
to orders as
\begin{equation}
L_0\, dt = e\vec A\cdot\vec{dR} - e\phi\,dt
\end{equation}
\begin{eqnarray}
L_1\,dt = e\vec r\cdot\grad\vec A\cdot\vec{dR} 
+ e\vec A\cdot\vec{dr} 
+ \zperp\cdot\vec{dR} 
\hskip 2 cm\nonumber\\ {}
- \LP e\vec r\cdot\grad\phi + m{U^2\over 2} 
+ {1\over 2m}\abs{\zperp}^2\RP\,dt
\end{eqnarray}
\begin{equation}
L_2\,dt = e\vec r\cdot\grad\vec A\cdot\vec{dr} 
+ \LP \zpl\bunit + \zperp\RP\cdot\vec{dr} 
- \half e\LP\vec r\vec r\dotdot\grad\grad\RP\phi\,dt
\end{equation}
where the dependence of $\vec A$ and $\phi$ and $\grad$
is now upon $\vec R$ after
the Taylor expansion.  Due to the factors of $m/e$ the spatial variation
of $\bunit$ enters one order lower than that of $\vec A$ so we do not
expand it.

The lowest-order one-form is $L_0\,dt$.
Varying $\vec R$ we find
\begin{equation}
e\LP \grad\vec A\cdot\vec{dR} - \vec{dR}\cdot\grad\vec A\RP = e\grad\phi\,dt
\end{equation}
The solution of this is
\begin{equation}
  \Rdot_0 = {1\over B^2}\bigg(
\grad\phi\cdot(\grad\vec A) - (\grad\vec A)\cdot\grad\phi \bigg)
  + \bunit(\bunit\cdot\Rdot)
\end{equation}
with order zero denoted by the subscript.
This solution describes the
lowest-order ExB drift.  We note that at this level the parallel
component $\bunit\cdot\Rdot$ is indeterminate.  Since the parallel
dynamics enters at the next order however we leave this component at
zero and specify
\begin{equation}
\vec u_0 \equiv \Rdot_0 = \vec u_E
\qquad \vec u_E \equiv {1\over B^2}\grad\phi\cdot\vec F
\qquad \vec F = \grad\vec A - (\grad\vec A)^T
\end{equation}
with superscript $T$ denoting the transpose.  This operation is how the
drift tensor $\vec F$ enters the problem.  Establishing $\vec u_0$ as
$\vec u_E$ is the main result of this step.

The next-order one-form is $L_1\,dt$.  We first subtract the total
differential
\begin{equation}
dS_1 = d\LP\vec r\cdot\vec A\RP = \vec {dr}\cdot\vec A 
+ \vec{dR}\cdot\grad\vec A\cdot\vec r
\end{equation}
where we note that $\vec A$ depends on $\vec R$ but is static.
The subtraction yields
\begin{equation}
L_1\,dt - dS_1
= e\vec r\cdot\vec F\cdot\vec{dR} 
+ \LP \zpl\bunit + \zperp\RP\cdot\vec{dR} + \cdots
\end{equation}
where we've written only the terms appearing as part of $\vec p$.
We now choose
$\vec r$.  Customary operations in the theory set $\vec r$ to cancel
the gyromotion 
$\vec w$ out of $\vec p$, leaving
$\vec u_0$ there as a background term.  In our case we choose $\vec r$
to cancel the entire $\zperp$, leaving only the parallel piece in the
canonical momentum, which is now just $e\vec A+\zpl\bunit$.  This sets
\begin{equation}
e\vec r\cdot\vec F + \zperp = 0
\label{eq:eqa1}
\end{equation}
and the solution is
\begin{equation}
\vec r = {1\over eB^2}\zperp\cdot\vec F
\label{eq:eqa2}
\end{equation}
where we arbitrarily choose $\bunit\cdot\vec r=0$.  This $\vec r$ is the
same gyro-drift radius as in the Lie-transform version of this model
\cite{Miyato09}.   We also find
\begin{equation}
e\vec r\cdot\grad\phi = -\vec u_E\cdot\zperp
\end{equation}
Discarding the total differential term, the
first-order one-form correction is
\begin{equation}
L_1\,dt - dS_1 = \zpl\bunit\cdot\vec{dR} 
- \LP m{U^2\over 2} + {1\over 2m}\abs{\zperp - m\vec u_E}^2
- m{v_E^2\over 2}\RP dt
\end{equation}
where using the evaluation of $\vec r\cdot\grad\phi$ we have completed
the squares on the last two terms.
This step is how $\vec u_0$, now $\vec u_E$, explicitly enters the
expression of $H$.

The result up to this point is  almost good enough to build the
model, since we have obtained the quadratic field term in $\phi$
necessary to build its field equation.  However, we haven't found any
constraints on $\zperp$ yet.  To do that we have to proceed to
the next order and consider the details of gyromotion as Littlejohn
did.  

At second order the Lagrangian correction is
\begin{equation}
L_2\,dt = \LP e\vec r\cdot\grad\vec A
+ \zpl\bunit + \zperp\RP\cdot\vec{dr} 
- \half e\LP\vec r\vec r\dotdot\grad\grad\RP\phi\,dt
\end{equation}
The first step is to use $\bunit\cdot\vec{dr}=0$ to strip the $\zpl$ term
and then subtract 
\begin{equation}
dS_2=d\LP\half e\vec r\cdot\grad\vec A\cdot\vec r\RP
= \half e\vec r\cdot\grad\vec A\cdot\vec{dr}
+ \half e\vec{dr}\cdot\grad\vec A\cdot\vec r
\end{equation}
with $d\vec A$ contributions an order smaller than those from the 
gyromotion.
This is to symmetrise the form with $\grad\vec A$, obtaining 
\begin{equation}
L_2\,dt - dS_2 
= \LP \half e\vec r\cdot\vec F + \zperp\RP\cdot\vec{dr} 
- \half e\LP\vec r\vec r\dotdot\grad\grad\RP\phi\,dt
\end{equation}
Using Eq.\ (\ref{eq:eqa1}), this can be re-cast as
\begin{equation}
L_2\,dt - dS_2 = \half \zperp\cdot\vec{dr} 
- \half e\LP\vec r\vec r\dotdot\grad\grad\RP\phi\,dt
\end{equation}
To evaluate the first term we examine the details of the gyromotion.

We set up an auxiliary basis $\vec e_1$ and $\vec e_2$ for the
plane perpendicular to $\bunit$, with restrictions and
coordinate sense
\begin{equation}
\vec e_1\cdot\vec e_2=0
\qquad
\vec e_{1,2}\cdot\bunit_2=0
\qquad
\vec e_1\cross\vec e_2\cdot\bunit=1
\end{equation}
giving the signs.
In the Hamiltonian through first order, $H_0+H_1$, we identify the
quantity
\begin{equation}
  \zperp - m\vec u_E \equiv m\vec w
\end{equation}
as the gyromotion velocity in the rest frame of the lowest order
velocity $\vec u_0$, which is $\vec u_E$.
We introduce the gyrophase angle $\vartheta$ hence covering the plane
with $\vec w$ expressed in terms of
$w$ and $\vartheta$.  Due to
the large $\Omega$ the fast part of $\vec{dr}$ is solely due to
$d(\vec w\cdot\vec F)$ with contributions due to $\vec u_0$ down an
order.  Contributions due to $\grad e_{1,2}$ will give formal 
gyrophase
invariance, but all the others due to the spatial variation of 
$\vec B$ are neglected.
The motion $(\Omega\,dt)$ is described locally as a geometric
circle with angle variation $d\vartheta$, with $\vec w$ 
the directed gyration velocity, of magnitude $w$,
and $\vartheta$ the gyrophase angle.
Using Eqs.\ (\ref{eq:eqa1},\ref{eq:eqa2}), we are
left with
$\vec w\cdot\bunit\cross\vec{dw}$ which is then worked through
as $\vec{dw}\cdot\vec w\cross\bunit$ according to
\begin{equation}
\vec w = -w\LP\vec e_1\sin\vartheta + \vec e_2\cos\vartheta\RP
\end{equation}
\begin{equation}
\vec{dw}=
-w\LP\vec e_1\cos\vartheta - \vec e_2\sin\vartheta\RP d\vartheta
-w\vec{dR}\cdot\LP\grad\vec e_1\sin\vartheta 
	+ \grad\vec e_2\cos\vartheta\RP
+{dw\over w}\vec w
\end{equation}
\begin{equation}
\vec w\cross\bunit = w\LP\vec e_2\sin\vartheta - \vec e_1\cos\vartheta\RP
\end{equation}
The sense of the motion for ions is clockwise, for $\bunit$ out of the
plane toward the observer, hence the signs.
We then use $\vec e_1\cdot\vec e_2=0$ hence
$\grad(\vec e_1\cdot\vec e_2)=0$ to find
\begin{equation}
\vec w\cdot\vec{dr} = \vec{dw}\cdot\vec w\cross\bunit = {w^2\over\Omega}
\LP d\vartheta - \vec{dR}\cdot\grad\vec e_1\cdot\vec e_2\RP
\end{equation}
This is the minimal description of gyromotion which preserves gyrophase
invariance through rotations 
$\vartheta = \vartheta' + \alpha(\vec R)$.
The gyromotion appears in the second order Lagrangian as
\begin{equation}
  L_2\,dt - dS_2 = {mw^2\over 2\Omega}
  \LP d\vartheta - \vec{dR}\cdot\grad\vec e_1\cdot\vec e_2\RP
  - \half e\LP\vec r\vec r\dotdot\grad\grad\RP\phi\,dt
  \label{eq:lagr2a}
\end{equation}
Averaging the gradient components in $\LP\vec r\vec
r\dotdot\grad\grad\RP$ over $\vartheta$,
by setting $\vec r\vec r\to (r^2/2)\vec g_\perp$, with $\vec g_\perp$
the perpendicular metric,
produces the main FLR correction, and we obtain
\begin{equation}
  L_2\,dt - dS_2 = 
  M\LP d\vartheta - \vec{dR}\cdot\grad\vec e_1\cdot\vec e_2\RP
  - e {r^2\over 4}\ddpp\phi\,dt
  \label{eq:lagr2b}
\end{equation}
where we have identified
\begin{equation}
  M = {mw^2\over 2\Omega}
  \label{eq:defm}
\end{equation}
as the conserved quantity multiplying $d\vartheta$.  Since
$\vartheta$-dependence has been eliminated everywhere else in $L_p$ we
now note that $M$ is a constant of the motion and suitable for use as
a coordinate.

The piece due to $\vec W=\grad\vec e_1\cdot\vec e_2$
is small but formally important
since the $d\vartheta$ piece by itself is not gyrophase invariant.
If $\vartheta = \vartheta' + \alpha(\vec R)$ then the combination
$d\vartheta - \vec W\cdot\vec{dR}$ is invariant (use the dependence of
$\vec e_{1,2}$ on $\vartheta$ and $d\alpha=\vec{dR}\cdot\grad\alpha$ to
show it).  In practice the gyromotion drops out of the kinetic equation
anyway, since $\ppvh{}$ and $dM/dt$ vanish.  The $M\vec W$ piece is
a small $O(r/L_B)^2$ correction to the $(r/L_B)$ drifts and does not
introduce any new charge-separation effects.  In practice no numerical
simulation to date keeps it.  For profile scales $\Lpp\ll L_B$
the higher-order terms from the contribution of $u_E^2$ to $r^2$ are
larger and these receive the attention.

We have now accounted for $w^2$ and $r^2$ appearing only through their
magnitudes so we combine the result $L_{0,1,2}\,dt$ as
\begin{equation}
L_p\,dt = \LP e\vec A+\zpl\bunit 
	-M\vec W\RP\cdot\vec{dR}
        +M\,d\vartheta - H\,dt
\end{equation}
with Hamiltonian
\begin{equation}
H = m{U^2\over 2}+M\Omega+\LP 1+{r^2\over 4}\ddpp\RP e\phi 
	- m{u_E^2\over 2}
\label{eq:ham}
\end{equation}
where
\begin{equation}
mU = \zpl - e\Apl 
\qquad
u_E^2 = {1\over B^2}\abs{\dpp\phi}^2
\qquad
r^2 = {2M\Omega + mu_E^2\over m\Omega^2}
\label{eq:ue2}
\end{equation}
where the extra differentials $dS_{1,2}$ are dropped (formally changing
the representation).
This can be shown (cf.\ Ref.\ \cite{pet15})
to be a low-$\kpp$ and low-$\beta$ version of the
result of Ref.\ \cite{Hahm88a}.

The importance of the second-order step is to establish $M$ as a
coordinate (the notation is from Hahm \cite{Hahm88}).  We may define a
magnetic moment $\mu$ as
\begin{equation}
  M\Omega \equiv \mu B \qquad\hbox{hence}\quad \mu = {e\over m}M
\end{equation}
Since $M$ is invariant, so is $\mu$.  It is interpreted as
the magnetic moment of the gyrocenter in the reference
frame co-moving with velocity $\vec u_E$.
Note that $M$ is first defined after noting in Eq.\ (\ref{eq:lagr2a})
that the combination $mw^2/2\Omega$ multiplies $d\vartheta$ and is
therefore invariant, hence re-writing as in Eq.\ (\ref{eq:lagr2b}), we
can replace the perpendicular energy term in $H$ with simply
$M\Omega$.  This is the step in which the fast gyromotion is decoupled
from the rest of the dynamics, since $\zperp$ no longer appears
explicitly. 

\subsection{The Field Lagrangian}

The field theory embeds this into a phase space
\begin{equation}
L = \sumsp\int\dL\,f\,L_p + \int\dV\,\scriptl_f
\end{equation}
where for shear-Alfv\'en conditions the field Lagrangian density is
\begin{equation}
\scriptl_f = \half\LP \eps_0 E^2-\mu_0^{-1}B^2\RP
\end{equation}
with $\vec B=\curl\vec A$ in entirety.  Under quasineutrality $\eps_0$
is taken to be small, and then under shear-Alfv\'en conditions the
field Lagrangian reduces to
\begin{equation}
  \scriptl_f = -{1\over 2\mu_0}\Bpp^2 \qquad\hbox{where}\quad
  \Bperp = \curl(\Apl\bunit)
\end{equation}
Only the perturbation due to $\Bperp$ appears in the field term, as
otherwise he full equilibrium current would have to be carried by the
gyrocenters.  We now have $\vec B = \curl\vec A$ as the background
magnetic field serving as an anchor, while $\Bperp$ carried by $\Apl$
accounts for the dynamics.

With these approximations we now re-write 
the system Lagrangian as
\begin{equation}
L = \sumsp\int\dL\,f\,L_p + \int\dV\,\scriptl_f
\label{eq:l}
\end{equation}
with gyrocenter and field Lagrangian pieces
\begin{equation}
L_p = \LP e\vec A+\zpl\bunit\RP\cdot\Rdot
	+ M\LP \thetadot - \vec W\cdot\Rdot\RP
        - H
\qquad
\scriptl_f = -{1\over 2\mu_0}\Bpp^2
\end{equation}
with Hamiltonian
\begin{equation}
H = m{U^2\over 2}+M\Omega+\LP 1+{r^2\over 4}\ddpp\RP e\phi 
	- m{u_E^2\over 2}
\label{eq:hh}
\end{equation}
and as auxiliaries the parallel velocity and the squares of the ExB
velocity and the gyro-drift radius
\begin{equation}
mU = \zpl - e\Apl 
\qquad
u_E^2 = {1\over B^2}\abs{\dpp\phi}^2
\qquad
r^2 = {2M\Omega + mu_E^2\over m\Omega^2}
\end{equation}
This is now a complete description of the dynamical system.
In following subsections, we derive the Euler-Lagrange equations for the
gyrocenters, the gyrokinetic equation for their distribution function,
and then the Euler-Lagrangian equations for the field variables giving
their self-consistent equations in the model.

\subsection{The Euler-Lagrange Equations for Gyrocenters}

Gyrocenter motion itself arises from $L_p$  only. We note that 
derivatives arise from variations with respect to the phase space 
coordinates holding each other fixed. Hence we note again that
the spatial gradient operator is taken with respect to gyrocenter
positions $\vec R$ holding $\zpl$ fixed. We define geometric quantities
\begin{equation}
e\Astar = e\vec A + \zpl\bunit - M\vec W
\qquad
\Bstar = \curl\Astar
\end{equation}
The derivatives of $H$ are
\begin{equation}
\grad H = e\grad\phi_E + M\grad\Omega_E - e U\grad\Apl
\end{equation}
spatially, and in the velocity space coordinates
\begin{equation}
\ppzpz{H} = mU
\qquad
\pMM{H} = \Omega_E
\qquad
\pvhvh{H} = 0
\end{equation}
Auxiliary quantities are
\begin{equation}
e\phi_E = e\phi - m{u_E^2\over 2}\LP 1-{\Omega_{exb}\over 2\Omega}\RP
\qquad
\Omega_E = \Omega + \half\Omega_{exb}
\qquad
\Omega_{exb} = {1\over B}\ddpp\phi
\end{equation}
Varying $\vec R$ and $\zpl$ together in $L_p\,dt$ and setting the
coefficients of the variation components to zero yields the drift motion
\begin{equation}
\Bpl\Rdot = {1\over e}\bunit\cross\grad H + \ppzpz{H}\Bstar
\qquad
\Bpl\pzdot = -\Bstar\cdot\grad H
\label{eq:driftmotion}
\end{equation}
and varying $\vartheta$ and $M$ yields the gyromotion
\begin{equation}
\Mdot=0
\qquad
\thetadot = \pMM{H} + \vec W\cdot\Rdot
\label{eq:gyromotion}
\end{equation}
In the drift motion the general form of the phase space volume element
is
\begin{equation}
\Bpl = e\pzz{\Astar}\cdot\Bstar
\label{eq:bpl}
\end{equation}
where in our case the $z$-coordinate is $\zpl$.

Since $\zpl$ enters $\Astar$ only through $\bunit$, we have
\begin{equation}
e\pzz{\Astar}=\bunit \qquad\hbox{hence}\quad \Bpl = \bunit\cdot\Bstar
\end{equation}
recovering the more usual expressions, also involving
$\bunit\cross\grad H$ in the drift motion.
In fact these forms are not general, since other choices for the
$z$-coordinate produce their own representations for $\ppz{\Astar}$
and hence the other expressions. But the forms in
Eqs.\ (\ref{eq:driftmotion},\ref{eq:bpl}) are general to any choice
determining the $z$-coordinate and $\Astar$.

\subsection{The Gyrokinetic Equation}

The operations to get to this are familiar.  We first observe that 
not only $\div\Bstar=0$ but also
\begin{equation}
{1\over e}\curl\LP e\pzz{\Astar} \RP - \pzz{\Bstar} = 0
\end{equation}
for any representation.   It follows that the
requirement of phase space incompressibility 
\begin{equation}
\div\Bpl\Rdot + \pzz{}\Bpl\pzdot = 0
\end{equation}
is satisfied.
The distribution function $f$ is just the density of
gyrocenters in phase space, so we have its continuity equation
\begin{equation}
\ptt{f} + \Rdot\cdot\grad f + \pzdot\pzz{f} = 0
\end{equation}
in advection form using the phase-space incompressibility.  There is no
$M$ term since $\Mdot=0$ and no $\vartheta$ term since $\ppvh{f}=0$.
This is
our application of Liouville's theorem.  In our case we started with
\begin{equation}
  z\to\zpl \qquad e\ppzpz{\Astar} = \bunit
\end{equation}
so that the results for the
gyrocenter motion lead to
\begin{equation}
\Bpl\ptt{f} + {1\over e}\bunit\cdot\grad H\cross\grad f
	+ \Bstar\cdot\LP\ppzpz{H}\grad f - \ppzpz{f}\grad H\RP = 0
\label{eq:gk}
\end{equation}
This is our gyrokinetic equation.  One thing to note is that the only
appearance of $M\vec W$ is in its contribution to $\Bstar$ and $\Bpl$
and it is small there and introduces no new effects.  This is why it is
usually dropped in computations.

\subsection{The Field Equations}

Given the system Lagrangian in Eq.\ (\ref{eq:l}) we find the equations
for $\phi$ and $\Apl$ by varying them in $L_p\,dt$ and setting the
coefficients of the variations (\ie, the functional derivatives) to
zero.

The induction equation for $\Apl$ is found by varying $L$ with respect
to $\Apl$,
\begin{equation}
\bunit\cdot\curl\LB\curl\LP\Apl\bunit\RP\RB = \mu_0\Jpl
\label{eq:ind}
\end{equation}
where
\begin{equation}
\Jpl = \sumsp\int\dW\, eU\, f
\end{equation}
is the gyrocenter current and arises from the appearance of $\Apl$ in
$H$. 
Eq.\ (\ref{eq:ind}) is the gyrokinetic Amp\`ere's law.

The polarisation equation for $\phi$ is found by varying $L$ with respect
to $\phi$,
\begin{equation}
\div \LP N_E\dpp\phi + \ddpp P_E\RP + \rhogc = 0
\label{eq:pol}
\end{equation}
where
\begin{equation}
\rhogc = \sumsp ne
\qquad
N_E = \sumsp {nm\over B^2}\LP 1-{\Omega_{exb}\over 2\Omega}\RP
\qquad
P_E = \sumsp {m\over 2eB^2}\LP p_\perp+nm{u_E^2\over 2}\RP
\end{equation}
are the gyrocenter charge density, the
polarisability, and the generalised FLR correction to the
gyrocenter charge density, with moment quantities given by
\begin{equation}
n = \int\dW\,f \qquad p_\perp = \int\dW\,M\Omega\,f
\end{equation}
These represent the gyrocenter density and perpendicular pressure.
Eq.\ (\ref{eq:pol}) is the gyrokinetic Poisson equation.
In this case all the contributions arise from $H$ solely.

Together, Eqs.\ (\ref{eq:gk},\ref{eq:ind},\ref{eq:pol}) describe the
complete self consistent dynamical system which arises from the
description due to the Lagrangian in Eq.\ (\ref{eq:l}).

\subsection{The Conventional Tokamak Case}

In a tokamak, the background magnetic field is axisymmetric.  The
general form is 
\begin{equation}
\vec B=\curl\vec A = I\grad\varphi + \grad\psi\cross\grad\varphi
\end{equation}
where $\varphi$ is the geometric toroidal angle about the symmetry
axis ($R=0$). In the conventional tokamak limit both the plasma beta
and the effective field line pitch angle away from purely toroidal are
small, so that we can re-cast the geometry as 
\begin{equation}
I = B_0R_0=\hbox{constant}
\qquad
\bunit = R\grad\varphi
\qquad
B = {I\over R}
\end{equation}
where $R$ is the toroidal major radius and $R_0$ and $B_0$ are
constants giving the reference values of $R$ and $B$.  This resuults
in
\begin{equation}
\Bperp = \curl\LP\Apl R\grad\varphi\RP \qquad \Bpp = {1\over
  R}\abs{\dpp(\Apl R)}
\end{equation}
in the field Lagrangian density.  If the full equilibrium current is
considered to be carried by the gyrocenters, then the field Lagrangian
density is generalised to 
\begin{equation}
\scriptl_f = -{1\over 2\mu_0 R^2}\abs{\dpp\LP\psi+\Apl R\RP}^2
\end{equation}
Hence $\Apl R$ acts as a perturbation to $\psi$ in a conventional
tokamak model.  Moreover, in this case
$\grad\varphi\cdot\curl(\zpl R\grad\varphi)$ vanishes when setting up
the coordinate Jacobian, so that $\Bpl$ reduces to $B$.

In the conventional tokamak case case the gyrokinetic Amp\`ere's law in
Eq.\ (\ref{eq:ind})
becomes
\begin{equation}
-R^2\div{1\over R^2}\dpp\LP\psi+\Apl R\RP = \mu_0 R\Jpl
\end{equation}
if $\psi$ is included in $\scriptl_f$.  
Using this form with $\psi$ requires the equilibrium current to be
contained in $\Jpl$.  If the latter is not, then $\psi$ is not
included in $\scriptl_f$ and then as a result does not appear in the
gyrokinetic Amp\`ere's law.

\section{General Phase Space Jacobian and the Four-Bracket Form}

We always assume the use of $M$ as one of the velocity-space
coordinates since it is a conserved quantity in the motion and
the gyromotion has little consequence for the rest of the dynamics.  But
the $z$-coordinate can be chosen differently (\eg, unperturbed energy
or angular momentum).  
To show the usefulness of the derivation method
we leave the $z$-coordinate general and
allow $\Astar$ to be a general function of all the coordinates.
The result is a generalised bracket form which is helpful for building
consistent computations.  We do assume $\Astar$ remains independent of
time so that canonical representation is maintained.

The gyrocenter Lagrangian is in general
\begin{equation}
L_p = e\Astar \cdot\Rdot + M\thetadot - H
\end{equation}
The gyromotion separates out in the same way as above.  We note that
$L_p$ can be re-written in terms of phase space four-vectors
\begin{equation}
L_p = e A^*_a \Zdot^a + M\thetadot - H
\end{equation}
with indices $a$ one of $\{ijk,z\}$ with $z$ the fourth coordinate.
The indices $\{ijk\}$ range over the spatial dimensions, and we will
use $\eps^{ijk}$ as the Levi-Civita pseudotensor of rank three with
units $g^{-1/2}$, the inverse spatial volume element given by
\begin{equation}
  {1\over\sqrt{g}} = \grad x^1\cross\grad x^2\cdot\grad x^3
\end{equation}
with the three spatial coordinates in order in a right handed system.
In general the component $A^*_z$ is zero but the $A^*_a$ can have
derivatives in any coordinate except $\vartheta$.  
Varying the differential action 
$L_p\,dt$ we have
\begin{equation}
\delta L_p\,dt = e \delta Z^a A^*_{b,a} dZ^b
+ e  A^*_b d(\delta Z^b) - \delta Z^a\,H_{,a}\,dt
+ \delta M(\cdots)
+ \delta\vartheta(\cdots)
\end{equation}
concentrating only on the four coordinates spanned by $\{a\}$ 
or $\{b\}$.  Subtracting the relevant differential we obtain
\begin{equation}
\delta L_p\,dt - e  d(A^*_b\,\delta Z^b)
= \delta Z^a \LB e\LP A^*_{b,a} - A^*_{a,b}\RP dZ^b - H_{,a}\,dt\RB
+ \delta M(\cdots)
+ \delta\vartheta(\cdots)
\end{equation}
This yields as Euler-Lagrange equations for the four coordinates 
$(\vec R,z)$
\begin{equation}
e\LP A^*_{p,a} - A^*_{a,p}\RP dZ^p = H_{,a}\,dt
\end{equation}
and re-labelling the existing summed index $b$ as $p$ for convenience
with what happens next.

We anticipate solving this by operating from the left with the 4D
Levi-Civita pseudotensor without units, denoted $\Epslash^{abcd}$, which
takes values $\pm 1$ or $0$ according to positive/negative perturbation
of indices from $\{1234\}$ or repeated indices.  Then, doubly
contracting with $A^*_{c,d}$ produces
\begin{equation}
e A^*_{c,d}\Epslash^{abcd}\LP A^*_{p,a} - A^*_{a,p}\RP dZ^p 
= \Epslash^{abcd}H_{,a}A^*_{c,d}\,dt
\end{equation}
In each step we will use the fact that there is no $A^*_z$ so that index
$z$ must be among the derivatives.

First considering index $b$ is $z$ so that $\{acd\}$ are the spatial
$\{ijk\}$ indices (the sign remains since it is two exchanges between
$ijkz$ and $izjk$), we have
\begin{equation}
e A^*_{j,k}\epslash^{ijk}\LP A^*_{p,i} - A^*_{i,p}\RP dZ^p 
= \epslash^{ijk}H_{,i}A^*_{j,k}\,dt
\end{equation}
where the lower case $\epslash^{ijk}$ is the 3D 
Levi-Civita pseudotensor without units.  We note that
\begin{equation}
\epslash^{ijk} = \sqrt{g}\eps^{ijk}
\qquad
\eps^{ijk}A^*_{j,k} = -(B^*)^i
\qquad
A^*_{l,i} - A^*_{i,l} = \eps_{ilm}(B^*)^m
\end{equation}
where $\{lm\}$ are also spatial indices.
We expand in terms of index $p$ being spatial or $z$, so that
\begin{equation}
e A^*_{j,k}\epslash^{ijk}\LB\LP A^*_{l,i} - A^*_{i,l}\RP dZ^l
- A^*_{i,z} dz\RB
= \epslash^{ijk}H_{,i}A^*_{j,k}\,dt
\end{equation}
The spatial terms cancel mutually since
\begin{equation}
\epslash^{ijk}A^*_{j,k} = -\sqrt{g}(B^*)^i
\qquad
A^*_{l,i} - A^*_{i,l} = \eps_{ilm}(B^*)^m
\end{equation}
so that the combination is proportional to $(B^*)^i\eps^{ilm}(B^*)^m=0$.
The last piece is
\begin{equation}
\sqrt{g}(B^*)^ie A^*_{i,z} \Zdot^z = \epslash^{ijk}H_{,i}A^*_{j,k}
\end{equation}
and here we define
\begin{equation}
\Bpl = e A^*_{i,z}(B^*)^i
\label{eq:bplz}
\end{equation}
as written in Eq.\ (\ref{eq:bpl}), so that
\begin{equation}
\Zdot^z = \Epsilon^{izjk}H_{,i}A^*_{j,k}
\end{equation}
now establishing the units of $\Epsilon^{abcd}$ to be
$(\sqrt{g}\Bpl)^{-1}$.

Next we choose index $b$ to be spatial so that one of the others is $z$,
but $c$ cannot be $z$, which leaves only $a$ and $d$ as choices.
We expand according to whether index $a$ is spatial or $z$ and the same
for index $p$, so that
\begin{equation}
e A^*_{j,k}\Epslash^{zijk}\LP A^*_{l,z}\RP dZ^l
+ e A^*_{k,z}\Epslash^{ijkz}
\LB\LP A^*_{l,i} - A^*_{i,l}\RP dZ^l - A^*_{i,z} dZ^z\RB
= \Epslash^{ijkz}H_{,i}A^*_{k,z}\,dt
\end{equation}
noting that any occurrence of $A^*_z$ drops out.  Noting that $\{zijk\}$
is 3 exchanges away from $\{ijkz\}$ we have
\begin{equation}
-e A^*_{j,k}\epslash^{ijk}\LP A^*_{l,z}\RP dZ^l
+ e A^*_{k,z}\epslash^{ijk}
\LB\LP A^*_{l,i} - A^*_{i,l}\RP dZ^l - A^*_{i,z} dZ^z\RB
= \Epslash^{ijkz}H_{,i}A^*_{k,z}\,dt
\end{equation}
The $A^*_{i,z}$ term drops due to antisymmetry of $\epslash^{ijk}$ leaving
\begin{equation}
e (B^*)^i\LP A^*_{l,z}\RP dZ^l 
+ e A^*_{k,z}\epslash^{ijk}\LB \eps_{ilm}(B^*)^m dZ^l\RB
= \Epslash^{ijkz}H_{,i}A^*_{k,z}\,dt
\end{equation}
Finally, contracting the 3D Levi-Civita tensors, we find
\begin{equation}
e A^*_{i,z}(B^*)^i dZ^j 
= {1\over\sqrt{g}}\Epslash^{ijkz}H_{,i}A^*_{k,z}\,dt
\end{equation}
Collectively we have proven
\begin{equation}
e A^*_{c,d}\Epslash^{abcd}\LP A^*_{p,a} - A^*_{a,p}\RP 
= \sqrt{g}\Bpl\delta^b{}_p
\end{equation}
with $\Bpl$ defined in Eq.\ (\ref{eq:bplz}) which is equivalent to
Eq.\ (\ref{eq:bpl}).  
The solution to the Euler-Lagrange equations is
\begin{equation}
\Zdot^b = \Epsilon^{abcd}H_{,a}A^*_{c,d}
\label{eq:zdot}
\end{equation}
The gyrokinetic equation is then
\begin{equation}
\ptt{f} + \Epsilon^{abcd}H_{,a}f_{,b}A^*_{c,d} = 0
\end{equation}
for any choice of coordinates under canonical representation
(this is Eq.\ 24 of Ref.\ \cite{momcons}).
In general, on the right hand side will be placed a collision operator
and perhaps external source and sink terms.  This form greatly
facilitates the construction of computational models since with a good
discretisation of a bracket structure (e.g., the Arakawa \cite{Arakawa} or
Morinishi \cite{Morinishi98} ones), the conservation properties of the
bracket are preserved and therefore the numerical scheme will be closely
conservative (this will depend on the timestep scheme).

\section{Other Choices for Phase Space Coordinates}

The usefulness of this version of the
Euler-Lagrange derivation is to show the generality of this definition
of $\Bpl$, which reduces to $\bunit\cdot\Bstar$ only if 
$e\pt\Astar/\pt z = \bunit$.  We first recover the standard forms
using $\vpl$ for the parallel phase-space coordinate,
for use as a guide.  Then,
we give two examples: 
an electrostatic model with use of
unperturbed energy $z=m\vpl^2/2+M\Omega$ together with $M$ as velocity
space coordinates (``energy representation''), 
and an electromagnetic model with use of
total canonical angular momentum $z=e(\psi+\Apl R)+m\vpl R$ 
together with $M$ as velocity space coordinates (``momentum
representation'').  The latter seems to be promising for studies of
gyrokinetic MHD.

\subsection{conventional representation}

The conventional representation uses $\vpl$ as the $z$-coordinate.  In the
electrostatic case, this is the same as using $\zpl$ as above since the
only difference is the factor of $m$ which is normalised away.  
Starting with
\begin{equation}
\Astar = \vec A + {m\over e}\vpl\bunit
\end{equation}
we obtain
\begin{equation}
{e\over m}{\pt\Astar\over\pt\vpl} = \bunit
\qquad
\Bstar = \vec B + {m\over e}\vpl\curl\bunit
\end{equation}
and then
\begin{equation}
\Bpl = \bunit\cdot\vec\Bstar = B + {m\over e}\vpl\bunit\cdot\curl\bunit
\end{equation}
which are what is usually given \cite{Hahm88}, also following the
conventional language of the drift-kinetic predecessor of gyrokinetic
theory \cite{Morozov66}.  Since the Hamiltonian $H$
has spatial gradients only through $\phi$ plus FLR
corrections and $M\Omega$, the ExB and grad-B drifts arise from the
spatial drifts in $\bunit\cross\grad H$, while the curvature drift arises
from the part due to $\curl\bunit$ (whose perpendicular
component is proportional to $\bunit\cdot\grad\bunit$, the actual
curvature) in $\Bstar$ where one factor of $m\vpl$ arises from $\ppz{H}$
and the other $\vpl$ from $\Bstar$ to produce the multiplier in
$m\vpl^2 \bunit\cdot\grad\bunit$ with which one is familiar.
Only if we keep $\Apl$
or ExB velocity corrections in the canonical momentum
\cite{Brizard95,Hahm96}, especially if time dependent,
does any of this significantly change.  But then canonical
representation is broken and we do not pursue that here.
We turn to the energy representation to highlight the effect on the
form of $\Bstar$ and $\Bpl$ and on the resulting equations.

\subsection{energy representation}

In an electrostatic
model with an energy representation the unperturbed energy $mv^2/2$ is
used as the $z$-coordinate, and 
the parallel velocity function,
Hamiltonian, and spatial canonical momentum are
\begin{equation}
m{U^2\over 2} = z - M\Omega
\qquad
H = z + e\phi_E
\qquad
e\Astar = e\vec A + mU\bunit - M\vec W
\end{equation}
with $\phi_E$ containing the FLR effects.  Then through the dependence
of $U$ upon $z$, we obtain
\begin{equation}
\Bpl = e\pzz{\Astar}\cdot\Bstar = {1\over U}\bunit\cdot\Bstar
\end{equation}
For the Euler-Lagrange equations we still have $\Mdot=0$ but
\begin{equation}
\thetadot = {eB\over mU}\bunit\cdot\Rdot + \vec W\cdot\Rdot
\end{equation}
awaits solution of the other dimensions.  Solving Eq.\ (\ref{eq:zdot})
for the rest we have
\begin{equation}
\Bpl\Rdot = \bunit\cross\grad\phi_E + U\Bstar
\qquad
\Bpl\zdot = -U\Bstar\cdot\grad(e\phi_E)
\end{equation}
(don't forget to apply $\Mdot=0$ in evaluating the $\delta\vec R$ terms)
and now the gyromotion agrees with Eq.\ (\ref{eq:gyromotion}) since
$\bunit\cdot\Rdot=U$ still holds.
We see that $\Bstar=\curl\Astar$
contains not only the curvature drift but also the
$\grad B$-drift.  This can be useful in an electrostatic model and
indeed such a model has been constructed with trapped-electron dynamics
in mind since under delta-f conditions both energy and $M$ are
constants of the motion \cite{Candy06}.
However,
the singularity at $U=0$ and the complications with $\Apl$ as part of
$\Astar$ make questionable the usefulness of the energy representation
in a total-f electromagnetic model.

\subsection{momentum representation}

Another choice is to combine the $\psi$ and $\Apl R$ in a
large-scale electromagnetic model neglecting $\vec W$
so that we re-define
\begin{equation}
z = e\LP\psi + \Apl R\RP + m\vpl R
\end{equation}
with the parallel velocity function becoming
\begin{equation}
mUR = z - e\LP\psi + \Apl R\RP 
\end{equation}
with $H$ remaining as in Eq.\ (\ref{eq:hh}) but with this version of $U$.
In this case
\begin{equation}
\Astar = \Apol + z\grad\varphi
\end{equation}
where $\Apol$ is the magnetic potential for $\Btor=I\grad\varphi$, the
toroidal magnetic field.
Now, the modified magnetic field $\Bstar$ is simply $\Btor$, since the
curl of $z\grad\varphi$ with $z$ held fixed is zero, and $\psi$ is part
of $z$.  We then have
\begin{equation}
e\pzz{\Astar} = \grad\varphi
\qquad\hbox{hence}\qquad
\Bpl = \Btor\cdot\grad\varphi = {I\over R^2}
\end{equation}
Since $z$ is the full toroidal canonical angular momentum it is
conserved except for $\pt H/\pt\varphi$.  In the equilibrium magnetic
field both $M$ and $z$ are conserved and the motion is purely
spatial.  Now, both the $\grad B$ and curvature drifts arise from $\grad
H$ since
\begin{equation}
\grad H = e\grad\phi_E + M\grad\Omega_E - mU^2\grad\log R
	- e {U\over R}\grad\LP\psi+\Apl R\RP
\end{equation}
in this representation.  In a non field-aligned global model we
can choose spatial coordinates $\{xy\varphi\}$ such that
\begin{equation}
x = \log{R\over R_0} \qquad y = {Z\over R_0} \qquad \sqrt{g}=R_0R^2
\end{equation}
The Euler-Lagrange equations for the gyrocenters are
\begin{equation}
\dtt{x} = -{1/e\over IR_0}\pyy{H} \qquad \dtt{y} = {1/e\over IR_0}\pxx{H}
\qquad
\dtt{\varphi} = \pzz{H} \qquad \dtt{z} = -\pvpvp{H}
\end{equation}
with $\sqrt{g}\Bpl = IR_0$ a constant.
In a 2D equilibrium relaxation model it is even simpler since
$\pt/\pt\varphi = 0$ and hence $dz/dt=0$, leaving
\begin{equation}
\ptt{f} + {1/e\over IR_0}[H,f]_{xy} = C(f)
\end{equation}
with a collision operator.  Together with the field equations (Eqs.\
\ref{eq:ind},\ref{eq:pol}) this would describe the 2D electromagnetic
gyrokinetic model.  It could be useful to computational studies of
equilibrium relaxation with an X-point (arising from contributions due
to external coils).

\subsection{section summary}

The main point of these examples is that where the various drifts arise
(from $H$ or from $\Bstar$) depends on the representation and no one of
them is more valid than another.  This pertains especially to the forms
of $\Bstar$ and $\Bpl$ so one has to examine the representation used to
be able to check whether the choices written down are consistent.
Ultimately this is a matter of knowing the starting point (choice of
coordinates and of the forms in $L\,dt$) so that the resulting equations
and geometric quantities are properly checked.

\section{MHD and Tokamak Equilibrium}

In general magnetohydrodynamics (MHD) the relevant low frequency
approximation is the combination 
$\beta\ll 1$ and $\omega\ll\kpp v_A$
in the ordering \cite{Strauss76,Strauss77}.
In the context of global tokamak MHD this was first cast as an aspect
ratio expansion: the ratio of $a$, the reference minor radius, to 
$R_0$, the reference major radius (usually the geometric axis), 
should satisfy $a/R_0\ll 1$.  The reasoning is
to keep shear-Alfv\'en dynamics, which have $\omega\sim v_A/qR_0$, but
not the compressional Alfv\'en dynamics, with $\omega\sim v_A/a$,
which requires $a/qR_0$ to be a small parameter.  Since in the Grad-Shafranov
equation $F_{dia}=RB_{tor}$ enters squared, the departures of $F_{dia}$
from $I={\rm constant}$ enter at $O(a/qR_0)^2$ for the current along
with $O(\beta)$ for the pressure.  These are at most one to four
percent in conventional tokamaks, which is why this ``Reduced
MHD'' treatment is in use.  The near-constancy of $F_{dia}$ is easily
verified in standard equilibrium computations of conventional tokamaks,
which can be defined as $a/R_0\sim 1/3$ and a $q$-profile giving values
in the range 1 to 2 at the magnetic axis and 3 to 10 at the edge.  This
means that the square of $\rho/qR$, with $\rho$ the minor radius
of a given flux surface with $q=q(\rho)$, is
small everywhere.  This is in line with keeping $\Apl$, but not $\Aperp$,
together with $\phi$ in the field variables.  Then, for tokamaks we
further take $F_{dia}=I$ and
$B=I/R$ and $\bunit\to R\grad\varphi$ in the Lagrangian.

In this illustration we use a Lagrangian which neglects the FLR
corrections and work in the $z=\zpl R$ representation.
The equilibrium $\psi$ is still kept separate from $\Apl R$, so that
$\vec B$ still includes the poloidal field due to $\psi$, 
but the current carried by $f$ is assumed to include the euqilibrium
piece, so that $\psi$ is included in the field Lagrangian density.
We have
\begin{equation}
L_p = e\Astar \cdot\Rdot + M\thetadot - H
\qquad
\scriptl_f = -{\Bpp^2\over 2\mu_0}
\end{equation}
with
\begin{equation}
e\Astar = e\vec A + z\grad\varphi
\qquad
H = m{U^2\over 2}+M\Omega+e\phi_E
\qquad
mU = {z\over R} - e\Apl
\end{equation}
\begin{equation}
\phi_E = \phi - {m\over e}{u_E^2\over 2}
\qquad
u_E^2={1\over B^2}\abs{\dpp\phi}^2
\qquad
\Bpp^2 = {1\over R^2}\abs{\dpp\LP\psi+\Apl R\RP}^2
\end{equation}
and
\begin{equation}
\Bstar = \vec B
\qquad
\Bpl = {B\over R} = {I\over R^2}
\end{equation}
The gyrokinetic equation then can be written
\begin{equation}
{B\over R}\ptt{f} + {1\over e}\grad\varphi\cross\grad H\cdot\grad f
+ \vec B\cdot\LP\pzz{H}\grad f-\pzz{f}\grad H\RP = 0
\end{equation}
The field equations are then
\begin{equation}
\vor = \sumsp\intW ef 
\qquad\hbox{where}\qquad 
\vor = - \div{\rho_m\over B^2}\dpp\phi
\label{eq:charge}
\end{equation}
\begin{equation}
-\Delta^*\LP\psi+\Apl R\RP = \mu_0 \Jpl R
\qquad\hbox{where}\qquad 
\Delta^* = R^2\div{1\over R^2}\dpp
\end{equation}
The relevant moment variables are
\begin{equation}
\rhom = \sumsp\intW mf
\qquad
\Jpl = \sumsp\intW eUf
\end{equation}
giving the mass density and the parallel current.

The gist of this is that the two main Reduced MHD equations are found
from the time derivatives of the two field equations, yielding the
vorticity equation and the Ohm's law, respectively.  These are
essentially gyrofluid moment equations, since the terms are evaluated
using velocity-space moments of the terms in the gyrokinetic equation.
This is facilitated by using the divergence form of the gyrokinetic
equation,
\begin{equation}
{B\over R}\ptt{f} 
+ \div\LB f {1\over e}\grad\varphi\cross\grad H\RB
- \pzz{}\LB f\vec B\cdot\grad H \RB = 0
\label{eq:gkz}
\end{equation}
and noting that $\int\dW/\Bpl$ commutes past $\grad$ and annihilates
$\ppz{}$.  The derivatives of $H$ are
\begin{equation}
\grad H = e\grad\phi_E - e {U\over R}\grad\LP\Apl R\RP
- \LP mU^2 + M\Omega\RP\grad\log R
\qqquad
\pzz{H} = {U\over R}
\end{equation}
We will be dealing with moments over unity and over $z$, in the
vorticity equation and Ohm's law, respectively.

The first equation to consider is charge conservation.  We take the time
derivative of Eq.\ (\ref{eq:charge}) and evaluate the $\ppt{f}$ terms,
to obtain
\begin{equation}
\ptt{\vor} + \LB\phi,\vor{R\over B}\RB 
	+ \LB{mu_E^2\over 2},n_i{R\over B}\RB
= B\dpl{\Jpl\over B} - \LB\log R^2,{\ppp+\Ppl\over 2}{R\over B}\RB
\label{eq:vor}
\end{equation}
where the spatial bracket structure denotes
\begin{equation}
[f,g] = \grad f\cross\grad g\cdot\grad\varphi
\end{equation}
and $\dpl$ combines the background magnetic field with the
perturbation according to
\begin{equation}
\vec B_T = I\grad\varphi + \grad\LP\psi+\Apl R\RP\cross\grad\varphi
\qqquad
B\dpl = \vec B_T\cdot\grad
\end{equation}
and the terms with $\log R$ are the generalised curvature terms.
Eq.\ (\ref{eq:vor}) is our generalisation of the Reduced MHD vorticity
equation.  The conventional form is recovered by neglecting finite-Mach
corrections and taking the pressure to be isotropic, leaving
\begin{equation}
\ptt{\vor} + \LB\phi,\vor{R\over B}\RB 
= B\dpl{\Jpl\over B} - \LB\log R^2,p{R\over B}\RB
\label{eq:vorrmhd}
\end{equation}
which is the Reduced MHD vorticity equation.

The second equation to consider is the one for the parallel electric
field.  We take the time derivative of Eq.\ (\ref{eq:ind}) and evaluate
the $\ppt{f}$ terms, to obtain
\begin{equation}
-R^2\div{1\over R^2}\dpp\LP R\ptt{\Apl}\RP = \mu_0 R
\sumsp\int\dW\, e\LP U\ptt{f} + f\ptt{U}\RP
\end{equation}
With $z=\zpl R$ we find as an auxiliary
\begin{equation}
\ptt{U} = -{e\over m}\ptt{\Apl}
\end{equation}
since $R$ is not time-dependent.
Pulling the $\Apl$ terms to the left side we arrive at
\begin{equation}
  \LP d_e^{-2}-R^2\div{1\over R^2}\dpp\RP\LP R\ptt{\Apl}\RP 
	= \mu_0 R\sumsp\int\dW\, {e\over m}\LP mU\ptt{f}\RP
\end{equation}
where the skin depth $d_e$ is given by
\begin{equation}
d_e^{-2} = \sumsp {\mu_0 ne^2\over m} \qqquad n = \int\dW\, f
\label{eq:wp}
\end{equation}
Each species contributes to both sides of this through inverse mass,
since the moments of $mU$ of the gyrokinetic equation scale like the
pressure gradient or the parallel electric field.  The latter comes from
the $\ppz{}$ term in Eq.\ (\ref{eq:gkz}) since $\ppz{(mUR)}=1$.  In the
MHD limit where $n_e e\grad\phi \gg \grad p$ for all species and the skin
depth $d_e$ is taken to be small along with $m_e$, the only
significant terms are the the electron contribution to $d_e$ and the
terms involving $\Bdel\phi$ and $[\Apl R,\phi]$ from the electron
contribution to $\ppt{f}$.  This leaves
\begin{equation}
\ptt{\Apl} + \dpl \phi = 0
\label{eq:aplrmhd}
\end{equation}
which is the Reduced MHD Ohm's law.
If collisional dissipation is added then the resistivity term adds to
the right hand side.

The equilibrium condition for the currents is found by setting the right
side of the 
Reduced MHD vorticity equation (Eq.\ \ref{eq:vorrmhd}) to zero.
Assuming $p=p(\psi)$ and $B=I/R$ the curvature term yields
\begin{equation}
  \LB\log R^2,p{R\over B}\RB 
  = cI\ppsps{p}\grad R^2\cross\grad\psi\cdot\grad\varphi
\end{equation}
Assuming that $\Apl$ and $\pt/\pt\varphi$ vanish, we also have
\begin{equation}
B\dpl\to \grad\psi\cross\grad\varphi\cdot\grad
\qquad\hbox{and}\qquad
\Delta^* \psi \to -\mu_0 \Jpl R
\end{equation}
and therefore
\begin{equation}
\grad\psi\cross\grad\varphi\cdot\grad\LB \Delta^*\psi + \mu_0\ppsps{p}R^2\RB=0
\label{eq:dplequil}
\end{equation}
which is the parallel gradient of the Grad-Shafranov equation
\begin{equation}
F_{dia}\ppsps{F_{dia}} + \Delta^*\psi + \mu_0 \ppsps{p}R^2=0
\label{eq:gradshaf}
\end{equation}
since the first term is a flux function (i.e., of $\psi$ only, like $p$).
In other words, the full tokamak equilibrium is recovered, since the
integral of Eq.\ (\ref{eq:dplequil}) produces an arbitrary flux
function which without loss of generality can be set such that
Eq.\ (\ref{eq:gradshaf}) is recovered.  This is the statement of
tokamak equilibrium under Reduced MHD.

These operations summarise the capture of Reduced MHD and the
Grad-Shafranov equilibrium by gyrokinetic theory \cite{Miyato09}.
The time derivative of the gyrokinetic Poisson equation yields the
Reduced MHD vorticity equation, and the time derivative of the
gyrokinetic Ampere's law yields the Reduced MHD Ohm's law.  The
equilibrium of the Reduced MHD vorticity equation in the absence of
strong flows yields the parallel gradient of the Grad-Shafranov
equation, or the Reduced MHD equilibrium.  This result is contained in
the more general treatment {\it On the gyrokinetic equilibrium}
by Qin \etal\ in Ref.\ \cite{Qin00}.

\section{Summary}

This outline has shown how the gyrokinetic representation of low
frequency plasma dynamics can be derived in a manner more transparent
than the ones which use Lie transforms and differential form calculus.
A small and straightforward re-casting of Littlejohn's gauge transform
method, originally used for the drift-kinetic case, is sufficient to
derive the gyrokinetic Lagrangian allowing for shear-Alfv\'en
electromagnetic electron responses and for finite-amplitude ExB flows.
If the flow amplitude is small, such that the ExB contribution to
gyroaveraging is negligible, and the plasma beta is low enough that
electromagnetic induction in electron responses is negligible, then
the model reduces to the conventional small-fluctuation, electrostatic
one that is most familiar.  The equations derived from the gyrokinetic
Lagrangian are not only the kinetic equation of motion, one per
species, but also the self consistent electric and shear-Alfv\'en
magnetic field potentials, to which each species contributes through
charge and current density.  The symmetry principles of a Lagrangian
field theory guarantee energetic consistency in what has just been
derived.  The equations derived from the gyrokinetic Lagrangian can
then serve as the basis for deriving the gyrofluid model within the
same representation which then has the same level of consistency.
These methods then form the basis for well-constructed numerical
models in computations of turbulence, MHD, energetic particle
dynamics, and equilibrium and transport in magnetised plasmas.  The
moment equations exhibiting the correspondences to MHD also indicate
the usefulness of this approach to generally energetically consistent
gyrofluid equations.

{

\input paper.bbl
}


\appendix

\section{On not gyroaveraging the magnetic potential}

In conventional gyrokinetic theory all of the time-dependent fields are
gyroaveraged in a form which can be represented as, e.g., $J_0\phi$ for
the electrostatic potential.  However, in most of my work the magnetic
potential $\Apl$ is not gyroaveraged.  The reason is that its dynamics
is controlled by the electrons with ion contributions entering through
the small mass ratio $m_e/M_i$.  Here we note that
$\rs^2=(M_i/m_e)\rho_e^2$, so that any ion-based FLR effect combines
with the mass ratio to produce a correction commensurate with an
electron FLR effect.  Unless the scale range 
$\rho_s^{-1}<\kpp<\rho_e^{-1}$ is explicitly treated, 
any global computation of mesoscale MHD plus drift Alfv\'en turbulence
will find this a negligible effect.

To see this we repeat the derivation of the Reduced MHD Ohm's law
in Eq.\ (\ref{eq:aplrmhd}).
The inverse skin depth defined in Eq.\ (\ref{eq:wp})
is clearly dominated by the electron term.  On the
right side the moment with $mU$ is taken of the gyrokinetic equation.  
All the terms are of similar size, with the main
ones being $ne\dpl\phi$ and $\dpl\Ppl$ with $\Ppl$ being the moment of
$mU^2$ over $f$.
Each species therefore contributes according to the multiplier 
$\mu_0 ne^2/m$ on the $\dpl\phi$ term which gives the static parallel electric
field, and according to the multiplier
$\mu_0 e/m$ on the pressure terms.  With all
pressures of similar size, the electron contributions are dominant here
too.  Hence, gyroaveraging $\Apl$ does not bring any significant
corrections and it is not necessary.  The main exception to this might
be global simulation of energetic particle MHD phenomena.  We leave such
things to future work.

\section{Ordering and Cancellations}

We have treated these before \cite{momcons}, but in the meantime there
is a version of the polarisation cancellation in the momentum
conservation law which is easier to see.  It is given here.

The toroidal canonical momentum conservation law in phase space is
\begin{equation}
\ptt{}\Pphi f + {1\over\sqrt{g}\Bpl}\pzzp{}\sqrt{g}\Bpl\Pphi f\zpdot
+ f\pvpvp{H} = 0
\end{equation}
where the specific toroidal canonical momentum is
\begin{equation}
\Pphi = \zpl\bphi + e\psi
\end{equation}
and $\bphi$ is the toroidal covariant unit vector component
(usually approximated as the toroidal major radius $R$),
and we are using the $z=\zpl$ representation in canonical structure
(cf.\ Eqs.\ 34,50 of Ref.\ \cite{momcons}).

The issue is the $\psi$ terms which appear with the factor of $ne$
and are of order unity.  With a neoclassical flow, the parallel momentum
itself is $O(\delta)$ in the small parameter $\delta=\rs/\Lpp$.  The
transport is given by a fluctuations in parallel momentum and ExB
velocity, each another order down so that the momentum flux is
$O(\delta^3)$.  Hence the belief that $O(\delta^3)$ drifts in 
$e\psi\,f\Rdot$ might be commensurate with this.

However, the largest terms in this equation cancel exactly, leaving the
toroidal ExB mometum term which in this context is $O(\delta^2)$.  Now
the largest term is the parallel momentum content at $O(\delta)$ and the
transport at $O(\delta^3)$ which is actually $O(\delta^2)$ relative to
the content. 
With the leading order eliminated each term is promoted by one order,
and the result is a transport equation in which the ordering is the same
as in any other transport equation: content at lowest $O(1)$, and
transport at $O(\delta^2)$.  It does not matter whether we call these
$O(\delta)$ and $O(\delta^3)$, since the relative order is the same.
This was already shown in Eqs.\ (73--80) of Ref.\ \cite{momcons}, but we
sketch it again here.

The charge conservation law is found by varying $\phi$ in the Lagrangian
to obtain the polarisation equation (Eq.\ \ref{eq:pol}), and then taking
the time derivative.  We may write these generally as
\begin{equation}
\div\vec P = \rhogc = \sumsp\int\dW\,ef
\qqquad
\div\ptt{\vec P} = \sumsp\int\dW\,e\ptt{f} = - \div\Jgc
\end{equation}
with the gyrocenter charge density and current defined as
\begin{equation}
\rhogc = \sumsp\int\dW\,ef
\qqquad
\Jgc = \sumsp\int\dW\,ef\Rdot
\end{equation}
We may always write $\rhogc$ as the divergence of a polarisation vector
$\vec P$ because all dependence of $L$ upon $\phi$ except for the first
$e\phi$ term in $H$ enters through gradients of $\phi$ (formally, $\vec
P$ is given by the functional derivative of $L$ with respect to $\vec E$ 
(cf.\ also Eqs.\ 15,16 of Ref.\ \cite{Morrison13}).
Note that the gyrocenter charge density is not zero and gyrocenter
dynamics is not ambipolar -- these concepts refer to the particle
density and dynamics.  Only under static conditions ($\ppt{}=0$) 
is the gyrocenter dynamics ambipolar, and only in the absence of an
electric field and FLR corrections does the gyrocenter charge density
vanish.  The charge conservation law is found by combining the
divergences into a total one,
\begin{equation}
\div\vec J = 0
\qqquad
\vec J = \Jpol+\Jgc
\qqquad
\Jpol = \ptt{\vec P}
\end{equation}
taking the time-dependent polarisation current $\Jpol$ into account.

Now considering the $\psi$ terms in the toroidal canonical momentum
conservation law, we have upon taking the velocity-space integral and
species sum
\begin{equation}
\sumsp\int\dW\ptt{}ef\psi 
+ \div\sumsp\int\dW\,ef\psi\Rdot
+ \cdots
\end{equation}
where the other terms are left to be treated separately.
Putting in for $\rhogc$ in the first term and $\Jgc$ in the second,
noting that $\psi$ pulls out of both the integral and the sum but not
the divergence, we have
\begin{equation}
\psi \ptt{\rhogc} + \div \psi\Jgc + \cdots
\end{equation}
Putting the polarisation current in for $\rhogc$ we have
\begin{equation}
\psi\div\Jpol + \div \psi\Jgc + \cdots
\end{equation}
In one term $\psi$ is inside the divergence, in the other not, so we do
the first one by parts, and combine the total divergences
\begin{equation}
-\Jpol\cdot\grad\psi + \div \LP\psi\vec J\RP + \cdots
\end{equation}
Since the divergence of the total current vanishes, this is re-cast as
\begin{equation}
-\Jpol\cdot\grad\psi + \vec J\cdot\grad\psi + \cdots
\end{equation}
Expressing the first term with the polarisation vector and noting
that $\psi$ is static, we have the exact result that
\begin{equation}
\psi \ptt{\rhogc} + \div \psi\Jgc 
= \ptt{}\LP -\vec P\cdot\grad\psi\RP
+ \vec J\cdot\grad\psi 
\end{equation}
Since $\psi$ is constant on a flux surface, it is proportional to $V$,
the volume enclosed by the surface.  Simply due to vector identities,
the flux surface average of the contraction between any
divergence-free vector and $\grad V$ vanishes \cite{Hazeltine73}.
So we take the flux surface average (denoted by angle brackets)
and bring the other terms back in,
we arrive at
\begin{equation}
\ptt{}\avg{\Mphi -\vec P\cdot\grad\psi}
+ {1\over V'}\ppsps{}V'\avg{\Piphi\cdot\grad\psi}
+ \avg{\sumsp\int\dW\,f\pvpvp{H}}
= 0
\end{equation}
where 
\begin{equation}
\Mphi = \sumsp\int\dW\,f\zpl\bphi
\qqquad
\Piphi = \sumsp\int\dW\,f\zpl\bphi\Rdot
\end{equation}
are the gyrocenter parallel momentum and parallel momentum flux,
the $f\pt H/\pt\varphi$ term gives the wave flux transport,
and $V'=dV/d\psi$ gives the dependence of the enclosed flux-surface
volume $V$ upon $\psi$.  This equation is the result of Section VII of
Ref.\ \cite{momcons}, with subsections A and B showing that the 
$f\pt H/\pt\varphi$ 
term is a total divergence.  Note that the treatments of the $\psi$
terms and of the $f\pt H/\pt\varphi$ term are separate and there are no
exchanges between the two.  In this equation, the two $f\zpl$ terms are
the largest ones and their relation as a transport equation is the same
as in those for particles or thermal energy.  This is the reason that
the lowest-order terms in $H$ due to the perturbed field variables
are the main ones for turbulent transport.  In rare circumstances the
Reynolds stress terms from $f\pt H/\pt\varphi$ can become concurrent but
are never dominant.  Lowest-order terms in $H$ always dominate in
the $\zpl$
terms, and second order terms in $H$ or $\scriptl_f$
always dominate in
the $\vec P\cdot\grad\psi$ and and $f\pt H/\pt\varphi$
terms. 

Similar considerations apply also to the charge conservation equation
itself, in which transport of $n$ by the part of $\Rdot$ due to $e\phi$
looks like a large term, but the same polarisation cancellation applies
to this term, so that the ExB velocity transports $\rhogc$ which is
replaced by the scalar quantity $\vor=-\div\vec P$ functioning as a
generalised vorticity.  The equation for $\vor$ is essentially a simple
generalisation of the vorticity equation in any two-fluid model.  This
has been shown elsewhere \cite{Lee01,Miyato09}.  Again, all the dominant
transport effects are given by the lowest order terms due to $\phi$ and
$\Apl$ in $H$ and the drift tensor due to $\Astar$, except for the
dependence of $\vor$ itself on the appearance of the ExB energy in $L$.

One can always check this by keeping some higher order terms in $H$ or
$\Astar$ in the theoretical model (for example, $mu_E^2$ compared to
$2M\Omega$ in $r^2$ in Eq.\ \ref{eq:ue2})
neglecting them {\it a priori} in computations, and then measuring
them {\it a posteriori} in the results.  A good example of how to do
this is given in some recent papers \cite{Idomura12,pet15}. 

\end{document}

%% file: symdefs.tex
\def\ie{{\it i.e.}}
\def\eg{{\it e.g.}}
\def\etal{{\it et al}}

\def\half{ {1\over 2} }

\def\eps{\epsilon}
 
\def\grapprox{\mathop{\lower.5ex \hbox{$\buildrel{\fivesy >}\over{\fivesy\sim}$}} \nolimits}
\def\lsapprox{\mathop{\lower.5ex \hbox{$\buildrel{\fivesy <}\over{\fivesy\sim}$}} \nolimits}
\def\grls{\mathop{\lower.5ex \hbox{$\buildrel{\fivesy >}\over{\fivesy <}$}} \nolimits}

\def\vec#1{{\bf #1}}

\def\avg#1{\left\langle #1 \right\rangle}
\def\abs#1{\left\vert #1 \right\vert}

\def\LB{\left\lbrack}
\def\RB{\right\rbrack}
\def\LP{\left (}
\def\RP{\right )}
\def\qqquad{\qquad\qquad}

\def\pt{\partial}

\def\pzz#1{{\partial #1\over\partial z}}
\def\pxx#1{{\partial #1\over\partial x}}
\def\pyy#1{{\partial #1\over\partial y}}

\def\ppsps#1{{\partial #1\over\partial \psi}}

\def\ptt#1{{\partial #1\over\partial t}}

\def\pvhvh#1{{\partial #1\over\partial \vartheta}}

\def\dtt#1{{d #1\over dt}}

\def\ppz#1{\partial #1/\partial z}

\def\ppt#1{\partial #1/\partial t}

\def\ppvh#1{\partial #1/\partial \vartheta}

\def\grad{\nabla}
\def\cross{{\bf \times}}
\def\div{\grad\cdot}

\def\curl{\grad\cross}
\def\dpl{\grad_\parallel}

\def\dpp{\grad_\perp}
\def\ddpp{\grad_\perp^2}

\def\bunit{\vec{b}}

\def\to{\rightarrow}

\def\dotdot{\!:\!}

\def\Bdel{\vec B\cdot\grad}

\def\Jpol{\vec J_p}

\def\vpp{v_\perp}
\def\vperp{\vec\vpp}

\def\Jpl{J_\parallel}

\def\Bperp{\vec B_\perp}
\def\Apl{A_\parallel}

\def\App{A_\perp}

\def\Aperp{\vec\App}

\def\vpl{v_\parallel}

\def\vor{\grad_\perp^2\phi}

\def\kpp{k_\perp}

\def\rs{\rho_s}

\def\Lpp{L_\perp}

\def\ptb{\widetilde}

\def\Afl{\ptb A_\parallel}

